\documentclass[11pt,letterpaper]{JHEP3}

\usepackage{epsfig,multicol,bbm}
\usepackage{amsfonts}
\usepackage{amsmath}
\usepackage{mqcommands}

\def\be{\begin{equation}}
\def\ee{\end{equation}}
\def\ba{\begin{eqnarray}}
\def\ea{\end{eqnarray}}
\def\beq{\begin{eqnarray}}
\def\eeq{\end{eqnarray}}

\newcommand{\pp}[1]{\mathcal P^{(#1)}}

\newcommand{\reef}[1]{(\ref{#1})}
\renewcommand{\eqref}[1]{(\ref{#1})}

\title{Massive Gravity Theories and limits of Ghost-free Bigravity models}

\author{Miguel F. Paulos$^{a}$, Andrew J. Tolley$^{b}$\\
$^a$ {\it Laboratoire de Physique Th\'eorique et Hautes Energies, CNRS UMR 7589,\\ Universit\'e Pierre et Marie Curie, 4 place Jussieu, 75252 Paris Cedex 05, France}\\
$^b$ {\it Department of Physics, Case Western Reserve University, 10900 Euclid Ave, Cleveland, OH 44106, USA}}

\abstract{
We construct a class of theories which extend New Massive Gravity to higher orders in curvature in any dimension. The lagrangians arise as limits of a new class of bimetric theories of Lovelock gravity, which are unitary theories free from the Boulware-Deser ghost. These Lovelock bigravity models represent the most general non-chiral ghost-free theories of an interacting massless and massive spin-two field in any dimension.
The scaling limit is taken in such a way that unitarity is explicitly broken, but the Boulware-Deser ghost remains absent. This automatically implies the existence of a holographic $c$-theorem for these theories. We also show that the Born-Infeld extension of New Massive Gravity falls into our class of models demonstrating that this theory is also free of the Boulware-Deser ghost. 
These results extend existing connections between New Massive Gravity, bigravity theories, Galileon theories and holographic $c$-theorems.

\vskip .5cm

{\rm E-mail:} \, {mpaulos@lpthe.jussieu.fr, andrew.j.tolley@case.edu}}


\begin{document}{\vskip 1cm}


\section{Introduction}

Recently there has been great interest in massive gravity models. On the one hand such models have been developed by modifying the Einstein-Hilbert action to include higher curvature terms, as exemplified by the three dimensional `New Massive Gravity' (NMG) model \cite{Bergshoeff:2009hq}. On the other hand there has been the recent successful definition of a nonlinear extension of the Fierz-Pauli mass term which is free from the Boulware-Deser instability \cite{Boulware:1973my} resulting in the first (and so far only) consistent Lorentz preserving theory of massive gravity in Minkowski spacetime for ($d >3$) - the de Rham-Gabadadze-Tolley model \cite{deRham:2010ik,deRham:2010kj}. In this article we shall demonstrate that both of these types of models can be seen as scaling limits of the ghost-free bigravity models of Hassan and Rosen, \cite{Hassan:2011zd} and use this observation to construct extensions of both types of massive gravity models which are guaranteed to be free from the  Boulware-Deser (BD) ghost, or instability \cite{Boulware:1973my}.

In building massive gravity models with higher curvatures, theories have been considered with quadratic \cite{Bergshoeff:2009hq}, cubic \cite{Sinha:2010ai} or arbitrary number of curvature corrections \cite{Paulos:2010ke},\cite{Gullu:2010pc}. By far the better studied theory is NMG, which arises at quadratic order in curvature. It is free from the Boulware-Deser ghost, and in asymptotically flat three-dimensional space, where the massless graviton is not propagating, an appropriate choice of sign of the Einstein-Hilbert term allows us to obtain a unitary theory of a massive spin-2 field.

One of the main guidelines in extending this theory was to demand the existence of a holographic $c$-function \cite{Myers:2010xs,Myers:2010tj}, in order to constrain the possible form of the higher curvature terms in the action. At the cubic level this is sufficient to completely fix the theory \cite{Sinha:2010ai}, but at higher orders this is not enough and degeneracies appear \cite{Paulos:2010ke}. Although these theories have the same spectrum as NMG (since they coincide with it at the linearized level), the absence or not of the BD instability is an important requirement which has not been verified for these theories thus far.

One of the most interesting extensions is the so-called DBI-gravity \cite{Gullu:2010pc}. This is a square-root action involving the metric and Einstein tensor, which also satisfies a holographic $c$-theorem \cite{Gullu:2010st}. A unitarity analysis of this theory has also been performed \cite{Gullu:2010em}, where it was found that this theory has the same properties at the quadratic level as NMG. Interestingly, it has also been found that such a theory can arise as a counterterm for gravity in $AdS_4$ \cite{Jatkar:2011ue,Sen:2012fc}, which in itself is enough reason to study it. 


In a separate development, there has been considerable effort to understand how to define a mass term for Einstein-Hilbert gravity which generalizes the Fierz-Pauli mass term without the introduction of the Boulware-Deser ghost. This culminated in the proposal of a completely nonlinear theory of massive gravity in four dimensions \cite{deRham:2010kj} which is guaranteed to be free of the Boulware-Deser ghost in a known decoupling limit \cite{deRham:2010ik} and has since been proven ghost-free nonlinearly \cite{Hassan:2011hr,Hassan:2011ea} following the arguments of \cite{deRham:2010kj}. An important realization made in \cite{Hassan:2011zd} was that the very same mass terms proposed in \cite{deRham:2010kj} could also be used to define ghost-free models of bigravity (bimetric theories). These theories include two metrics and because of the absence of the BD ghost describe consistent interacting theories of a single massless and a single massive spin-two field. By taking a well understood scaling limit which decouples the massless spin-two mode, these theories can be show to give rise to massive gravity on Minkowski  \cite{deRham:2010kj} or other fixed backgrounds \cite{Hassan:2011tf}.

Remarkably all of these theories can in appropriate decoupling limits be related to the Galileon models \cite{Nicolis:2008in}. To be precise, the dynamics of the helicity zero mode of the massive graviton are well approximated in an appropriate limit by a scalar field with a nonlinearly realized Galilean symmetry. This was shown for NMG in \cite{deRham:2011ca}, for de Rham-Gabadadze-Tolley massive gravity in \cite{deRham:2010gu,deRham:2010ik,deRham:2010kj} and  is equally applicable to the appropriate decoupling limit of all the models of massive gravity described in this article. 

In this paper we explore interconnections between these seemingly disparate subjects. We begin by showing in section \ref{limitsection} that NMG can be seen as a particular limit of the ghost-free bimetric theories of gravity of Hassan and Rosen. Generalizing the latter to include Lovelock terms and taking the very same limit, leads us to consider a large class of theories describing Lovelock gravity coupled to an auxiliary field with polynomial interactions. These theories are one of the main results in this note and are written down explicitly in equation \reef{actions}. They constitute higher curvature generalizations of NMG in any dimension, and which by construction are free from the BD instability. Further, in section \ref{determinant} we show that for a particular choice of parameters these theories are equivalent to the Born-Infeld gravity theory in $d=3$. In higher dimensions they provide novel determinant-type actions for gravity\footnote{However, see the very recent pre-print \cite{Yi:2012gh}.}.

In section \ref{BD instability} we do an explicit analysis of the dynamical content of our theories theories and show that they generically describe propagating massless and massive spin-2 modes. Further we show that the absence of the BD ghost arises due to important factors: firstly, all interaction terms take a Galileon form; secondly the coupling of the auxiliary field to the metric is made via a conserved tensor - in our case, a linear combination of the Lovelock tensors.

Finally we consider a large class of theories of gravity whose higher curvature terms depend only on the Schouten tensor. These can be thought of a slice through higher dimensional generalizations of the theories of \cite{Paulos:2010ke}, which includes the cubic gravity model first introduced in \cite{Sinha:2010ai}. By rewriting these theories in terms of an auxiliary field, and using the results of section \ref{BD instability} we show that generically these theories are not expected to be free from the Boulware-Deser ghost.
We finish this note with a short discussion.

\section{Scaling limits of bigravity theories}
\label{limitsection}

\subsection{Ghost-Free bigravity models in $d$-dimensions}
\label{bigravity}

We will be interested in a class of ghost-free $d$-dimensional bigravity models which are natural generalizations of the models introduced in \cite{Hassan:2011zd} utilizing the $d$-dimensional generalization of the finite number of allowed mass-terms given in \cite{deRham:2010kj}. These models describe two independent metrics interacting through a potential which depends on the characteristic polynomials of a certain matrix, and have the further properties that they are explicitly free from the Boulware-Deser instability and contain a single propagating massless spin-two and single massive-spin two modes.  We start by introducing the characteristic polynomials, and then describe the lagrangian of the models themselves.

The characteristic polynomials of a matrix $M$ are symmetric polynomials in the eigenvalues of that matrix. Alternatively, they are defined by
	\bea
	\pp N\equiv \delta^{a_1\ldots a_n}_{[b_1 \ldots b_n]}\, M_{a_1}^{b_1}\,\ldots M_{a_n}^{b_n}
	\eea
with the generalized Kronecker delta,
	\bea
	\delta^{a_1\ldots a_n}_{[b_1 \ldots b_n]}\equiv \frac{1}{n!}(\delta^{a_1}_{b_1}\ldots \delta^{a_n}_{b_n}+\mbox{perms}).
	\eea
Some properties of these polymomials are given in appendix \ref{poly}. For now we merely note that the first few polynomials are
	\bea
	\pp0(M_a^b)=1, \qquad \pp1(M_a^b)=[M], \qquad \pp2(M_a^b)=\frac 12 \left([M]^2-[M^2]\right) \nonumber \\
	\pp3(M_a^b)= \frac 16 \left([M]^3-3\, [M][M^2]+2\, [M^3]\right)
	\eea
with $[M]=M_a^a,\, [M^2]=M_{ab}\,M^{ba},\, [M^3]=M_{ab}M^{bc}M_{c}^a$, and so on.

We are now ready to introduce the bigravity theories. The simplest realization of these theories describe two dynamical metrics $g_{ab}$ and $f_{ab}$ with action  \cite{Hassan:2011zd}
\be
S = \int d^d x \left[ \frac{M_a^{d-2}}{2} \sqrt{-g} R[g]+ \frac{M_b^{d-2}}{2}\sqrt{-f} R[f] + \bar M^{d-2} m^2 \sqrt{-g} \, \mathcal U(g,f) \right] \label{bimetriclag}
\ee
where $\bar{M}^{-(d-2)}= M_a^{-(d-2)}+M_b^{-(d-2)}$
and the potential can be expressed in terms of the form  (these allowed mass terms were first derived for massive gravity in \cite{deRham:2010kj})
\begin{equation}
\label{eq:fullU}
\mathcal{U}(g,f)=\sum_{n=0}^d \alpha_n\pp n(\mathcal{K}).
\end{equation}
The $\alpha_n$ are free parameters, and
\be
\mathcal{K}^{a}_{b}(g,f)\equiv \delta^{a}_{b}-\sqrt{g^{a c}f_{c b}}.
\ee
More precisely $\mathcal{K}_{ab}=g_{a c }\mathcal{K}^{c}_{b}$ is the symmetric tensor defined in \cite{deRham:2010kj} that has the property 
\be
2 \mathcal{K}^{a}_{b}-\mathcal{K}^{a}_{c}\mathcal{K}^{c}_{b}=\delta^{a}_{b}-g^{a c}f_{c b}.
\ee
As shown in \cite{deRham:2010kj}, when $f_{ab}$ is the Minkowski metric $\mathcal{K}^{a}_{b}$ is the unique tensor that picks out the Galilean invariant combination of the helicity zero mode $\mathcal{K}^a_{b} \propto \partial^{a}\partial_b \pi$ in the appropriate decoupling limit $M_{a}, M_b \rightarrow \infty$ keeping $m^2 M_{a}$ fixed. This gives a simple explanation for why the mass term must be a characteristic polynomial in $\mathcal{K}_{ab}$ since only these combinations of $\partial_a \partial_b \pi $ are total derivatives, a necessary requirement to ensure that the equations of motion for $\pi$ are second order. It also explains the connection between decoupling limits of massive gravity models and Galileon models since the helicity-zero mode always enters in this Galileon invariant combination. 

It is a straightforward observation that it is always possible to take a scaling or decoupling limit of \reef{bimetriclag} for which $M_b \rightarrow \infty $, keeping $M_a$ fixed, that decouples the dynamics of the metric $f_{\mu\nu}=\bar{f}_{\mu\nu}+ M_b^{-(d-2)/2} \delta f_{\mu \nu}$ such that $\bar{f}_{\mu\nu}$ is essentially a fixed background metric \cite{Hassan:2011zd}. In this limit, the effective theory for $g_{\mu\nu}$ is a massive gravity theory (with the massless mode encapsuated by $\delta f_{\mu\nu}$ decoupled) defined on a fixed background metric $\bar{f}_{\mu\nu}$. In the special case for which $\bar{f}_{\mu\nu}=\eta_{\mu\nu}$ we reproduce the de Rham-Gabadadze-Tolley model \cite{deRham:2010kj}.

Although the coefficients $\alpha_n$ are arbitrary, we can without loss of generality set $\alpha_1=0$ (by conformally rescaling $f_{ab}$ and simultaneously shifting $\mathcal{K}$ by a multiple of the identity matrix) and $\alpha_2=1$ (by rescaling the mass). 
For the special choice of coefficients where $\alpha_n=\alpha^n $ for any $n$, we have
\bea
\mathcal{U}(g,H)=\mbox{det} \left(\delta_a^b+\alpha \, \mathcal K_a^b\right)
\eea
and so the generic case can be viewed as a deformation of this determinant. In the deformed determinant representation, it is easy to show that the action is symmetric under the interchange $g_{ab} \rightarrow f_{ab}$ with an appropriate redefinition of all the constants $\alpha_n$ in the action \cite{Hassan:2011vm}.  This follows from the determinant identity 
\be
\sqrt{-\det{g}} \, \det(\delta^{a}{}_{b}+\alpha \,  \mathcal{K}^{a}{}_{b}) = \alpha^d \sqrt{-\det{f}} \, \det(\delta^{a}{}_{b}+\alpha^{-1} \,  \mathcal{\tilde{K}}^{a}{}_{b}) ,
\ee
where 
\be
\tilde{\mathcal{K}}^{a}_{b}=\delta^{a}_{b}-\sqrt{f^{a \alpha}g_{\alpha b}}.
\ee

Although \cite{Hassan:2011zd} originally considered bigravity actions containing only the Einstein-Hilbert term, in higher dimensions one could equally well generalize this by considering each metric theory to be described by a Lovelock theory of gravity \cite{Lovelock1971}:

\bea
S_{\mbox{\tiny LL}} &=& \int d^d x \left[ \frac{M_a^{d-2}}{2} \sqrt{-g} \left(\sum c^{(K)}_{g} \mathcal L^{(K)}[g]\right)\right.\nonumber \\
&&\left.+ \frac{M_b^{d-2}}{2}\sqrt{-f} \left(\sum c^{(K)}_{f} \mathcal L^{(K)}[f]\right) + \bar M^{d-2} m^2 \sqrt{-g} \, \mathcal U(g,f) \right] \, ,
\label{lbigravity}
\eea
where $ \mathcal L^{(K)}$ denote the Lovelock invariants.
These `Lovelock bigravity' models are the most general non-chiral ghost-free theories of a single massless spin-two field coupled to a single massive spin-two field in arbitrary dimensions\footnote{This does not include the so-called Topologically Massive Gravity models \cite{TMG1,TMG2} - however these models treat left moving and right moving gravitons differently and should thus be distinguished from the classes of models discussed here. A recent alternative approach to constructing theories of massive gravity in a dual formulation has been proposed in \cite{Bergshoeff:2012ud} - however as yet there is no proof of the absence of BD ghost in these models}.
A short review of Lovelock theories of gravity is given in appendix \reef{llappendix}.  Suffice to say that the Lovelock lagrangians are the most general lagrangians leading to ghost-free metric theories with two derivative equations of motion. A generalization of the proof of the absence of a BD ghost for these Lovelock bigravity models is given in appendix  \reef{lbgappendix}. 

\subsection{NMG as a scaling limit}
\label{limit}

We now shall take a scaling limit of the actions \reef{bimetriclag} in which the Planck masses tend to infinity. However we shall take this limit in an unusual way, in that one of the masses will become infinitely negative. In this way we are explicitly breaking unitarity of the theory in dimensions $d>3$ but not for for the special case $d=3$. Define $f_{ab} = g_{ab}+ \lambda \, q_{ab}$. As it stands this is just a change of variables. However we will now consider what happens in the limit $\lambda \rightarrow 0$ combined with $M_a \rightarrow \infty$ with $M_a^{d-2}+M_b^{d-2}=M_P^{d-2}$ and $\lambda M_b^{d-2}\equiv A$ held fixed. We can without loss of generality choose to rescale $q_{ab}$ so that $A=M_P^{d-2}$. In this limit the kinetic part of the lagrangian becomes
\ba
{\cal L}_{\rm kinetic}&&  =\left[ \frac{M_a^{d-2}+M_b^{d-2}}{2} \sqrt{-g} R[g] -\frac{M_b^{d-2}}2 \, \lambda \,  \sqrt{-g}  q^{ab}G_{a b} + \cal O(\lambda)\right] \nonumber \\
&&\rightarrow \frac{M_P^{d-2}}2 \sqrt{-g}\left( R[g] - q^{ab}G_{a b} + \cal O(\lambda)\right)
\ea
Now let us look at the potential term. We note that in this limit
\be
\mathcal{K}_{a b} = - \frac{1}{2} \lambda \, q_{ab} + {\cal O}(\lambda^2) 
\ee
Thus we have
	\bea
	\pp n(\mathcal K) &=& \left(-\frac{\lambda}2\right)^n \pp n(q)
	\eea
We can choose to take the scaling limit such that
	\bea
	\bar{M}^{d-2} m^2 \alpha_n(- \lambda/2)^n \to \frac{1}{2}M_P^{d-2}\, \beta_n
	\eea
with by definition $\beta_0= -2 \Lambda$, the cosmological constant.
In this limit, the action becomes
	\bea
	S=\frac{M_P^{d-2}}2 \int \ud^d x \sqrt{-g}\left(R+2\Lambda-\,  q^{ab} G_{ab}+\sum_{n=2}^{d} \beta_n \pp n(q) \right) \label{auxBiMetric}
	\eea
Setting $\beta_n=0$ for $n>2$  we recover the auxiliary field formulation of NMG, and therefore we have shown that NMG arises as a particular limit of the bimetric theory of gravity. 

We can now repeat the logic with the general class of bimetric theories of Lovelock gravity, to find a new class of NMG type theories which are manifestly free from the BD instability. These models are described by actions of the form
	\bea
	S=\frac{M_P^{d-2}}2 \int \ud^d x \sqrt{-g}\left[\left(\sum_K a_K \mathcal L^{(K)}\right)+q^{ab}\left(\sum_K b_K  G^{(K)}_{ab}\right)+\sum_{n=2}^{d} \beta_n \pp n(q) \right] \label{actions}
	\eea
Here $G^{(K)}_{ab}$ are the Lovelock tensors defined in appendix \ref{llappendix} (essentially the equations of motion associated with a given Lovelock invariant). By construction, these theories generically describe two sets of propagating modes, a massless and a massive spin-2 mode, of which one is necessarily a ghost. 

For general $\beta_n$, integrating out the auxiliary field generates an infinite set of higher order curvature terms. For particular choices however we can write down the resulting lagrangian exactly in closed form. A simple example is given in appendix \reef{4der}. Another important class of examples correspond to picking the couplings such that the interaction terms become a determinant, and it is to this case that we now focus our attention.

\section{The determinant case}
\label{determinant}
Let us consider a particular case of the action \reef{auxBiMetric}, where we take
	\bea
	\beta_n=\alpha \gamma^{n-1}.
	\eea
and we consider the particular Lovelock theory which is Einstein gravity. In this case we can write
	\bea
	S &=& \frac{M_P^{d-2}}2 \int \ud^d x \sqrt{-g}\left(R+2\Lambda-q^{ab} G_{ab}+\frac{\alpha}{\gamma} \sum_{n=2}^{d} \gamma^n \pp n(q) \right)  \nonumber \\
	&=& \frac{M_P^{d-2}}2 \int \ud^d x \sqrt{-g}\left[R+2\hat \Lambda-q^{ab}\left(G_{ab}+\alpha g_{ab}\right) +\frac{\alpha}{\gamma} \, \mbox{det}(\delta_a^b+\gamma q_a^b) \right]
	\eea
with $\hat \Lambda=\Lambda-\frac{\alpha}{2\gamma} $. With this particular choice of coefficients, we will be able to solve for the auxiliary field. Indeed, its equation of motion is given by
	\bea
	G_{ab}+\alpha g_{ab}=\alpha \, \mbox{det}(\delta_a^b+\gamma q_a^b)\, \left[\left({1}+\gamma q\right)^{-1} \right]_{ab}
	\eea
Using this it is straightforward to show
	\bea
	\mbox{det}(\delta_a^b+\gamma q_a^b)&=&\left[\mbox{det}\left (\delta_a^b+\frac{1}{\alpha }\, G_a^b\right )\right]^{\frac 1{d-1}} \nonumber \\
	q^{ab}\left (G_{ab}+\alpha \, g_{ab}\right)&=& d\, \frac{\alpha}{\gamma}  \,
	\left[\mbox{det}\left (\delta_a^b+\frac{1}{\alpha}\, G_a^b\right )\right]^{\frac 1{d-1}}+\frac{d-2}{\gamma}\, R
	\eea
and therefore the action becomes
	\bea
	S &=& \frac{M_P^{d-2}}2 \int \ud^d x \sqrt{-g}\left[
	\left(1-\frac{d-2}{\gamma}\right)\, R+2\hat \Lambda-(d-1)\,\frac{\alpha}{\gamma}  \left[\, \mbox{det}\left (\delta_a^b+\frac{G_a^b}{\alpha}\right )\right]^{\frac 1{d-1}}
	\right]
	\eea
For $d=3$ this precisely becomes the action for Born-Infeld massive gravity \cite{Gullu:2010pc},\cite{Jatkar:2011ue} with an additional Einstein-Hilbert contribution. Furthermore for the special choice $\gamma=(d-2)$ we can reduce the action entirely to Born-Infeld NMG. 
For general $d$, and setting $\gamma=d-2$ we get
	\bea
	S &=& \frac{M_P^{d-2}}2 \int \ud^d x\left \{ \left(\sqrt{-g}\right)^{\frac{d-3}{d-1}}
	\,
	\frac{(d-1)}{(d-2)}\, \alpha \left[-\, \mbox{det}\left (g_{ab}-\frac{1}{\alpha} G_{ab}\right )\right]^{\frac 1{d-1}}
	+2 \sqrt{-g} \hat \Lambda
	\right\}
	\eea
For $d\neq 3$ this does not coincide with the higher dimensional generalizations of the Born-Infeld action, and appears to be a new class of theories. By construction, they are explicitly from from the BD ghost. It is also clear that natural generalizations of this class of models arise by considering the general Lovelock lagrangian to begin with. This leads e.g. to models of the form
	\bea
	S &=& \frac{M_P^{d-2}}2 \int \ud^d x\left \{ \left(\sqrt{-g}\right)^{\frac{d-3}{d-1}}
 \left[-\, \mbox{det}\left (\sum a_K G^{(K)}_{ab}\right )\right]^{\frac 1{d-1}}
	+2 \sqrt{-g} \Lambda
	\right\}
	\eea
which have yet to appear in the literature.

\section{Proof of absence of Boulware-Deser instability}
\label{BD instability}


In this section we will generalize the proof given in \cite{deRham:2011ca} of the absence of the Boulware-Deser ghost to the larger class of theories \reef{actions}.
Although this is guaranteed from their bimetric origin, it is instructive to see this in detail. We start by rewriting the action \reef{auxaction} as one with additional symmetries. We introduce new degrees of freedom by the replacement
\be
q_{a b}= Q_{a b} + 2\, \nabla_{(a}A_{b)}+ 2  \nabla_{a} \nabla_{b} \pi. \label{decomp}
\ee
The interpretation of these new degrees of freedom is that $\pi$ represents the helicity zero mode of the massive spin-two field and $A_a$ represents the helicity one mode.
Inevitably this decomposition introduces an additional linearized diffeomorphism symmetry $A_{a} \rightarrow A_{a}+ \chi_{a}$ combined with $Q_{ab} \rightarrow Q_{ab}-\nabla_{a} \chi_{b}-\nabla_{b} \chi_{a}$ , as well as an additional $U(1)$ symmetry $A_{a} \rightarrow A_{a}+ \nabla_{a} \chi$ together with $\pi \rightarrow \pi - \chi$. It is sometimes convenient work with the $U(1)$ gauge invariant combination $V_{a} = A_{a}+ \nabla_{a} \pi$.

The point of this decomposition is that the total number of degrees of freedom is 
\bea
{\rm d.o.f.}&=&2\times \frac{d(d+1)}2 \, (g_{ab}, Q_{a b})+ d \, (A_{a}) + 1 \, (\pi) \nonumber \\
&& - 2 \times \left(d \, \text{Diff}+d \, \text{L. Diff}+1 \, U(1) \right) = (d-1)^2-2 \label{dof}
\eea
which is the correct number of degrees of freedom for a massless and a massive spin-two fields in $d$ dimensions. 
However, this statement is only guaranteed if we can prove that after this substitution, the equations of motion for all the variables, in particular $\pi$ and the metric are second order. Unfortunately, this does not seem to be the case for the actions defined in \reef{auxaction} but it is the case for all actions of the form \reef{actions}.

We substitute the decomposition \reef{decomp} into the action. Consider first the interaction terms. These take the form
	\bea
	&&\pp N\equiv \delta^{a_1\ldots a_n}_{[b_1 \ldots b_n]}\, q_{a_1}^{b_1}\,\ldots q_{a_n}^{b_n}\nonumber \\
	&& \to \delta^{a_1\ldots a_n}_{[b_1 \ldots b_n]} \left( \nabla_{a_1} A^{b_1}+\nabla^{b_1} A^{a_1}+\nabla_{a_1}\nabla^{b_1} \pi\right)\ldots \left( \nabla_{a_n} A^{b_n}+\nabla^{b_n} A^{a_n}+\nabla_{a_n}\nabla^{b_n} \pi\right)
	\eea
It is clear that the equations of motion following from these terms cannot contain more than two derivatives acting on a single field, due to the totally antisymmetric nature of the generalized Kronecker delta. For instance, consider the term only involving the scalar $\pi$. The contribution to the equation of motion from such a term will be
	\bea
	\delta^{a_1 a_2\ldots a_n}_{[b_1 b_2 \ldots b_n]} \nabla_{a_1}\nabla^{b_1}\, \left(\nabla^{b_1}\nabla_{a_1}\pi\, \ldots \nabla^{b_n}\nabla_{a_n}\pi\right)
	\eea
Antisymmetric products of derivatives can be traded for curvatures, and hence the above does not contain terms with more than two derivatives acting on $\pi$. The same argument holds when the $A_a$ vector is included. That is because even with the vector included  $\pp N$ are at most linear in $\partial_t^2 \pi$ because of the antisymmetry of the generalized Kronecker delta ensures that the total number of time derivatives in any term is maximum two, and the latter fact is sufficient to guarantee that the contributions to the equations of motion for $\pi$ coming from these terms are second order in time-derivatives.

We also have to consider the contribution from the terms linear in $q$, which take the form:
	\be
	G^{(K)}_{ab}(2\nabla^{(a}A^{b)}+ 2  \nabla^{a}\nabla^{b} \pi)  = 2 \nabla^{a}(G^{(K)}_{ab} V^{b})-2 \nabla^{a}(G^{(K)}_{ab}) V^{b}.
	\ee
The first term on the RHS is a total derivative and doesn't contribute to the equations of motion. The second term would generically contribute, but since the Lovelock tensors $G^{(K)}_{ab}$ are conserved, this term is also vanishing. This is the crucial point on which our whole analysis rests: the linear $q_{ab}$ term must couple to a conserved tensor. Finally since the terms $G^{(K)}_{ab}$ are the equations of motion associated with a given Lovelock invariant, they are at most linear in $\partial_t^2 g_{ij}$ with no time derivatives of the metric in the coefficients and this is in turn sufficient to guarantee that the equations of motion obtained by varying with respect to the metric are second order.

In summary, it is the antisymmetry of the generalized Kronecker delta symbols in both the mass terms and the Lovelock tensor and the existence of $2d+1$ symmetries in this formulation that guarantees that all the equations are at most second order in time derivatives, and we conclude that the action indeed describes coupled massive and massless spin-2 modes and the Boulware-Deser ghost is absent.

Finally as explained in \cite{deRham:2011ca} in the form it is easy to take a decoupling limit which focuses on the interaction of $\pi$. Because of the manner in which $\pi$ is introduced, only in the two derivative combination $\nabla_a \nabla_b \pi$, it follows that in the decoupling limit $\pi$ exhibits a Galileon symmetry $\pi \rightarrow \pi + c_\mu x^{\mu}$ under which the combination $\partial_a \partial_b \pi$ is invariant. Thus the form of the interactions for $\pi$ in the decoupling limit (defined around Minkowski spacetime) are after appropriate diagonalization equivalent to the Galileon type interactions already well known from massive gravity \cite{deRham:2010ik,deRham:2010kj} and the Galileon \cite{Nicolis:2008in}.  

\section{Analysis of other New Massive gravity extensions}
\label{nmgext}
The class of theories we will be interested in in this section are defined in terms of characteristic polynomials of the Schouten tensor $S_{ab}$:
	\bea
	S_{ab}\equiv\frac{1}{(d-2)}\left( R_{ab}-\frac 1{2(d-1)} g_{ab} R\right).
	\eea
The class of theories we will consider are given by
	\bea
	S=\frac{M_P^{d-2}}2 \int \ud^d x \sqrt{-g}\,\left(R+2\Lambda+\sum_{n=0}^{d} \alpha_n \pp{n}(S_a^b)\right) \label{badactions}
	\eea
For $d=3$ this includes the $3d$ cubic extension of New Massive gravity given in \cite{Sinha:2010ai}. These theories satisfy a holographic $c$-theorem. To see this, we note that the Schouten tensor is related to the difference between the Weyl and Riemann tensors:
\bea
C_{abcd}-R_{abcd}=2\left(g_{a[c}\, S_{d] b}+g_{b[d}\, S_{c] a}\right)
\eea
This result implies that conformally flat solutions of Lovelock gravity can coincide with solutions of the above theories. For instance in $d=5$, the theory with all $\alpha_n$ set to zero except $\lambda_2$ (i.e. five-dimensional New Massive Gravity) will have the same conformally flat solutions as Gauss-Bonnet gravity. In particular, domain wall backgrounds given by
	\be
	\ud s^2=\ud r^2+ e^{2 A(r)}\ud \mbf x^2
	\ee
will take the same form in both theories. This immediately implies that NMG in $d=5$ satisfies a holographic $c$-theorem, and not only that, but the Euler anomaly will be the same as for Gauss-Bonnet gravity.

The action \reef{badactions} can be rewritten with the help of an auxiliary field $q_{ab}$:
	\bea
	S=\frac{M_P^{d-2}}2\,\int \ud^d x \sqrt{-g}\left(R-2\Lambda+\sum_{n=0}^d \frac{\alpha_n}{n-1} \left(q^{ab} \mathcal P_{ab}^{(n)}(S)-\pp n(q)\right) \label{auxaction}
	\right)
	\eea
with
	\bea
	\mathcal P_{ab}^{(n)}(M)\equiv \frac{\ud \pp n(M)}{\ud M^{ab}}.
	\eea
Now, for instance, if all $\alpha_n$ are zero except for $n=2$ we obtain the auxiliary field representation of $d$-dimensional new massive gravity,
	\bea
	S_{NMG}=\frac{M_P^{d-2}}2\,\int \ud^d x \sqrt{-g}\left(R-2\Lambda-\frac{\lambda_2}{(d-2)} q^{ab} G_{ab}+\frac{\lambda_2}{2} \left(q_{ab}q^{ab}-q^2 \right)
	\right)
	\eea
In any case, it is very easy to integrate out $f_{ab}$ to recover the original action \reef{badactions}, since its equation of motion implies $f_{ab}=S_{ab}$.

The main difference between \reef{auxaction} and the actions \reef{actions} is that the term which is linear in $q^{ab}$ now couples to a tensor which is not generically conserved, apart from the special case of New Massive gravity. Following the arguments in section \reef{BD instability}, we therefore do not expect this class of theories to be free from the Boulware-Deser instability because the equation of motion for the metric includes higher than two time derivatives. This applies to the model first proposed in \cite{Sinha:2010ai}, and although we shall not show it here in detail, also to the large class of models proposed in \cite{Paulos:2010ke}.

\section{Discussion}

In this note we have introduced a class of theories which extend the new massive gravity models. These action are written down in \ref{actions}, and they describe theories of propagating massless and massive spin-2 fields which free from the BD instability. The absence of the BD ghost is intimately tied to the special form of the interactions - the same form that arises in Galileon models \cite{Nicolis:2008in}, models of massive gravity in Minkowski space-time \cite{deRham:2010ik,deRham:2010kj} and bigravity models  \cite{Hassan:2011zd}. We have shown how the recently constructed unitary bimetric theories of gravity of Hassan and Rosen \cite{Hassan:2011ea,Hassan:2011hr}, can be generalized to the Lovelock bigravity models \reef{lbigravity} and secretly contain within them both the New Massive Gravity lagrangian, as well that of DBI gravity, at least in three-dimensions in an appropriately defined scaling limit. Of course this limit requires picking the wrong sign for one of the fields, explicitly breaking unitarity in the process. However in 3 dimensions unitarity can be maintained by absorbing the wrong sign terms into the non-dynamical massless graviton. For higher dimensionality we've shown how one can obtain other kinds of determinant lagrangians, which to the best of our knowledge have not appeared in the literature. It would be interesting to explore these theories further.

Our models constitute several parameter extensions of NMG in higher dimensions. At the linearized level they are identical to NMG, and as such have the same critical points. However, also by construction, our models are free from the BD instability at the nonlinearized level. This is an important point, as one of the things we might be interested in is taking the massless limit of the theory. In this limit, the two graviton modes become degenerate, and we expect logarithmic modes to appear. The theory becomes ``critical'' \cite{Bergshoeff:2010iy,Alishahiha:2011yb,Lu:2011zk}, and is expected to describe a logarithmic CFT \cite{Grumiller:2009sn,Skenderis:2009nt}. 
At higher orders, if we want the massless limit to be smooth we need absence of the BD ghost, something our models achieve explicitly. 

One of the results in this note is that seemingly the large class of models \reef{badactions} is {\em not} free from the BD ghost, even though they satisfy a holographic $c$-theorem. To understand this, notice that one way to diagnose the possible presence of the BD ghost is to look at the trace or conformal part of the metric. In particular, absence of the BD ghost implies that conformally flat metrics satisfy two derivative equations of motion. However, the reverse is not true in general, since absence of the BD ghost places constraints on the trace part of the metric even in non-conformally flat backgrounds. This is reflected in the results of sections \ref{BD instability} and \ref{nmgext}. In any case, this simple remark clarifies the connections found between massive gravity theories and holographic $c$-theorems \cite{Sinha:2010ai,Paulos:2010ke}. Indeed, the existence of a holographic $c$-function only demands that the equation of motion for a conformally flat metric mode should be two derivative, and therefore for theories where the BD ghost is absent, such as those in \reef{actions}, this is automatically guaranteed. 

\acknowledgments

AJT would like to thank Claudia de Rham for useful discussions and the Universit\'e de  Gen\`eve for hospitality whilst this work was being completed. MFP thanks Aninda Sinha for discussions and acknowledges funding from the LPTHE, Universite' Pierre et Marie Curie.

\appendix 	
\section{Characteristic polynomials}
\label{poly}
The tensor polynomials we are interested in take the form
	\bea
	\pp N\equiv \delta^{a_1\ldots a_n}_{[b_1 \ldots b_n]}\, M_{a_1}^{b_1}\,\ldots M_{a_n}^{b_n}
	\eea
with the generalized Kronecker delta,
	\bea
	\delta^{a_1\ldots a_n}_{[b_1 \ldots b_n]}\equiv \frac{1}{n!}(\delta^{a_1}_{b_1}\ldots \delta^{a_n}_{b_n}+\mbox{perms}).
	\eea
Define $\mathcal M_1\equiv M_a^{\ a}, \mathcal M_2\equiv \frac 12 M_{ab}M^{ba}, \mathcal M_3\equiv \frac 13 M_{a b}M^{b c} M_{c a}$, etc. Then we can write the polynomials in closed form:
	\bea
	\pp N(M)=\sum_{p(N)} \frac{(-1)^{N+L(p)}}{S}\prod_{i=1}^{L(p)} \mathcal M_{n_i}
	\eea
Some comments are required. Here $p(N)$ stands for integer partitions of $N$. Any such partition can be represented as a Young tableau, with each row representing an integer. Then $L(p)$ is nothing but the number of rows of the Young tableau, and the integers $n_i(p)$ stand for the number of boxes in the $i-th$ row of partition $p$. For instance, the partition
$	
	8=3+2+2+1
$%
 has $n_1=3, n_2=n_3=2, n_3=1$, and $L(p)=4$. Finally $S$ is the symmetry factor of the partition. An integer appearing $m$ times in a given partition contributes a factor $m!$ to it. In the example above the symmetry factor is $S=1\times 2! \times 1$. 

Equivalently, we can label partitions by integers $d_n$ which characterize the number of times integer $n$ appears. In this case we can write
	\bea
	\pp N(M)=(-1)^N\sum_{\{d_m\}}
	\prod_{n=1}^{+\infty} \frac{(-\mathcal M_n)^{d_n}}{d_n!}
	\eea
where the sum runs over all possible $d_m, m=1,\ldots +\infty$, subject to the constraint
	\bea
	\sum_{m=1}^{+\infty}\, m\, d_m=N
	\eea
Let us focus on the case where we have $N\times N$ matrices.
We then have:
	\bea
	\mbox{det}\,\left(M_a^{b}\right)=N! \, \pp N(M)
	\eea
We are interested in the case where $M_a^b=\delta_a^b+S_a^b$. Using the fact that
	\bea
	\delta^{a_1\ldots a_n}_{b_1 \ldots b_n}\delta_{a_1}^{b_1}\ldots \delta_{a_m}^{b_m}
	=\frac{(N-n+m)!}{(N-n)!}\frac{(n-m)!}{n!}\, \delta^{a_{m+1}\ldots a_n}_{b_{m+1}\ldots b_n}
	\eea
it is easy to show
	\bea
	\pp N(\delta +M)= \sum_{m=0}^N \pp m(M)
	\eea
	
\section{Lovelock theories of gravity}
\label{llappendix}
Lovelock theories of gravity are the most general second order gravity theories which are also free of ghosts when expanding around an arbitrary background and propagate only the degrees of freedom of a massless spin-two field \cite{Lovelock1971,Zumino1986}. Using the generalized Kronecker delta
	\bea
	\delta_{[b_1\ldots b_n]}^{a_1 \ldots a_n}\equiv \delta_{b_1}^{a_1}\ldots \delta_{b_n}^{a_n}\pm \mbox{permutations}
	\eea
we can write the $K$-th Lovelock invariant as
	\bea
	\mathcal L^{(K)}\equiv 
	\delta_{[c_1 d_1 \ldots c_K d_K]}^{a_1 b_1 \ldots a_K b_K}\, R_{a_1 b_1}^{\ \ \ \ c_1 d_1} \ldots R_{a_K b_K}^{\ \ \ \ c_K d_K} 
	\eea
Similarly the $K$-th Lovelock tensor is given by
	\bea
	\left(G^{(K)}\right)_e^f \equiv -\frac 12\,\delta_{[e\, c_1 d_1 \ldots c_K d_K]}^{f\, a_1 b_1 \ldots a_K b_K}\, R_{a_1 b_1}^{\ \ \ \ c_1 d_1} \ldots R_{a_K b_K}^{\ \ \ \ c_K d_K}
	\eea
The Lovelock tensor is nothing but the equation of motion following from a lagrangian given by the corresponding Lovelock invariant, which guarantees that its a covariantly conserved tensor, i.e.
	\bea
	\nabla_a \left(G^{(K)}\right)^{ab}=0
	\eea

The first few of these quantities are
	\bea
	\mathcal L^{(0)}&=&1, \qquad \mathcal L^{(1)}=R \nonumber \\
	\mathcal L^{(2)}&=&R_{abcd}R^{abcd}-4 R_{ab}R^{ab}+R^2 \nonumber \\
	G^{(0)}_{ab} &=&-\frac{g_{ab}}2 \qquad G^{(1)}_{ab} =R_{ab}-\frac 12 g_{ab} R \nonumber \\
	G^{(2)}_{ab} &=&2 R_{ab} R\,-4 R_{a c}R_b^{\ c}-4 R_{acbd}R^{cd}+2\, R_{acde}R_{b}^{\ cde}-\frac 12 g_{ab}\, \mathcal L^{(2)}
	\eea
Notice that by construction, both $\mathcal L^{(K)}$ and $G^{(K)}_{ab}$ are identically zero when $d<2K$. For $d=2K$ the invariant $\mathcal L^{(K)}$ is topological and therefore $G^{(K)}_{ab}$ is identically zero.

Finally, a useful consideration is that by definition we have
	\bea
	G^{(K)}_{ab}=\frac 12 \left( \frac{\partial \mathcal L^{(K)}}{\partial g^{ab}}-g_{ab}\, \mathcal L^{(K)}\right) 	\Rightarrow g^{ab} G^{(K)}_{ab}=\left(\frac{2K-D}2\right)\, \mathcal L^{(K)}
	\eea
where the partial derivative wrt the metric is defined keeping the Riemann tensor $R_{abcd}$ fixed, with this precise index structure.

\section{Lovelock bigravity}
\label{lbgappendix}

We shall now give a proof of the absence of a Boulware-Deser ghost applicable to the Lovelock bigravity models \reef{lbigravity}.
Consider the starting action 
\bea
S_{\mbox{\tiny LL}} &=& \int d^d x \left[ \frac{M_a^{d-2}}{2} \sqrt{-g} \left(\sum c^{(K)}_{g} \mathcal L^{(K)}[g]\right)\right.\nonumber \\
&&\left.+ \frac{M_b^{d-2}}{2}\sqrt{-f} \left(\sum c^{(K)}_{f} \mathcal L^{(K)}[f]\right) + \bar M^{d-2} m^2 \sqrt{-g} \, \mathcal U(g,f) \right]
\eea
Following \cite{Hassan:2011zd} we perform a double ADM decomposition of the two metrics. Associated with the metric $g_{\mu\nu}$ are the lapse $N$ and shift $N_i$. Associated with the metric $f_{\mu\nu}$ are the lapse $L$ and shift $L_i$. It is well-known that the Lovelock terms are the unique set of terms, in the appropriate dimension, that preserve the same number of degrees of freedom as Einstein gravity. An equivalent way of saying this is that the Lovelock terms are natural extensions to the action for Einstein gravity that do not change the phase space structure and have the same number of symmetries. As such, they can in principle be written in an identical canonical form. Given this fact the above action when written in canonical form is
\be
S_{\mbox{\tiny LL}} = \int d^d x \left[ \pi^{ij} \partial_t g_{ij} + p^{ij} \partial_t f_{ij} - N H -N_i H^i - L R - L_i R^i + {\cal L}_M \right] \, ,
\ee
where ${\cal L}_M$ is the mass-term
\be
{\cal L}_M =  \bar M^{d-2} m^2 \sqrt{-g} \, \mathcal U(g,f) \, ,
\ee
$H$, $H_i$ are the terms that in the absence of the mass term would give the Hamiltonian and momentum constraint for the metric $g_{\mu\nu}$, and $R$ and $R_i$ are the associated terms for the metric $f_{\mu\nu}$. In practice however this procedure is problematic because the reexpression of the time derivatives of the metric in terms of the conjugate momentum is impossible to perform explicitly because of the complicated nonlinear expression they take. This is because, even after integration by parts to remove all double time derivative terms, the Lovelock action will contain non-removable functions of $\partial_t \partial_k g_{ij}$ and similar terms for $f_{ij}$. In this case the correct definition of the conjugate momenta $\pi^{ij},p^{ij}$ are 
\be
\pi^{ij} = \frac{\partial {\cal L}}{\partial \partial_t g_{ij}} - \partial_k \frac{\partial {\cal L}}{\partial (\partial_t \partial_k g_{ij})} \, , 
\ee
and
\be
p^{ij} = \frac{\partial {\cal L}}{\partial \partial_t f_{ij}} - \partial_k \frac{\partial {\cal L}}{\partial (\partial_t \partial_k f_{ij})} \, ,
\ee
and here ${\cal L}$ is the total Lagrangian including factor of $\sqrt{-g}$.
To construct the canonical form of the action these equations must be inverted to give $\partial_t g_{ij}$ and $\partial_t f_{ij}$ in terms of $\pi^{ij}$ and $p^{ij}$. In general the solution is not unique, and the issue of different branches of solutions arises. For discussions on this issue see \cite{Teitelboim:1987zz} and \cite{Deser:2011zk}. However these complications do not affect the rest of the argument as we shall see. Locally the inversion may be performed within a given branch and the fact this inversion can be done in principle and that the above action would take the above form is all that counts in terms of establishing the correct number of degrees of freedom.

Since the Lovelock bigravity action \reef{lbigravity} is naturally invariant under a single copy of the diffeomorphism group in which $g_{\mu\nu}$ and $f_{\mu\nu}$ transform in the same way, we are guaranteed that the Hamiltonian will maintain $d$ first class constraints even in the presence of the mass term. These constraints arise because of the $2d$ non-dynamical fields $N,N_i, L, L_i$, there are $d$ linear combinations for which the mass term is linear in these combinations. An easier way to state this is that if we construct the $2d \times 2d$ matrix defined in the $2d$ dimensional space of non-dynamical degrees of freedom $N_0 = N, N_i=N_i, N_d=L, N_{d+i}=L_i, i=1,2, \dots, (d-1), A=0,1, \dots (2d-1)$ (see \cite{deRham:2011rn} for a related discussion)
\be
L_{AB} =  \frac{\partial^2 {\cal L}}{\partial N_A  \partial N_B}=  \frac{\partial^2 {\cal L}_M}{\partial N_A  \partial N_B} \, ,
\ee
then diffeomorphism invariance guarantees that the matrix $L_{AB} $ has at least $d$ zero eigenvalues associated with the existence of $d$ first class constraints in $d$ dimensions.
If it were the case that $L_{AB}$ had only $d$ zero eigenvalues, then the total number of phase space degrees of freedom would be $4 \times d(d-1)/2 - 2 \times d = 2 \times (d^2-2d) $ phase space or $(d-1)^2-1$ configuration space degrees of freedom. However the correct total number of degrees of freedom for a massless and massive spin-two field in $d$ dimensions is $(d-1)^2-2$. Thus the absence of the BD ghost implies that as usual there must exist two additional constraints, a primary one and a secondary one. The existence of an additional primary constraint implies that the matrix $L_{AB} $ must in fact have $d+1$ zero eigenvalues for the theory to be free of the BD ghost. However $L_{AB} $ depends only on the mass-term $\cal{L}_M$ and is hence independent of the precise form of $H, H_i ,L$ and $L_i$. In other words the question of whether there exists an additional primary constraint to remove the BD ghost is independent of the precise form of the massless limit of the theory. 
Thus the proof given in \cite{Hassan:2011zd} for Einstein-Hilbert gravity with a mass term already guarantees that $L_{AB} $ for the general Lovelock bigravity theory has $d+1$ zero eigenvalues, corresponding to the required $d$ first class and additional second class constraint. In fact the methodology of the proof is simply to perform the linear redefinition of variables that explicitly performs the diagonalization which makes the zero eigenvalues manifest. This additional second class constraint is sufficient to remove the BD instability. A complete analysis should demonstrate that a secondary constraint also arises following the argument of \cite{Hassan:2011ea}. We leave this to a future work to demonstrate, however let us note that as always the existence of a primary constraint is alone sufficient to negate the BD ghost argument. Furthermore the proof given in section \ref{BD instability} guarantees both primary and secondary constraints arise for all the new massive gravity extensions.

\section{An interesting four-derivative theory of gravity}
\label{4der}

In this section we consider the simplest non-trivial example of the general class of theories introduced in the main text. We consider the action written in terms of an auxiliary field:
	\bea
	S=\int \ud^d x \sqrt{-g}\left[R+2\Lambda-f^{ab} G^{(K)}_{ab}+m^2\left(f^{ab}f_{ab}-f^2\right)\right]
	\eea
We integrate out the auxiliary field $f_{ab}$ via its equation of motion which sets
	\bea
	f_{ab}=\frac{1}{4 m^2}\, \left[\frac{\partial \mathcal L^{(K)}}{\partial g^{ab}}-g_{ab}\left(\frac{2K-1}{d-1}\right)\, \mathcal L^{(K)}\right]
	\eea
The action becomes
	\bea
	S=\int \ud^d x \sqrt{-g}\left\{R+2\Lambda-\frac{1}{16 m^2}\left[\frac{\partial \mathcal L^{(K)}}{\partial g^{ab}}\frac{\partial \mathcal L^{(K)}}{\partial g_{ab}}-
	\left(\frac{d+4K(K-1)}{d-1}\right)
	\,\left(\mathcal L^{(K)}\right)^2\right]\right\}
	\eea
In the simple case $K=1$ we obtain
	\bea
	S=S_{\mbox{\tiny NMG}}=\int \ud^d x \sqrt{-g}\left[R+2\Lambda-\frac{1}{4 m^2}\left(R_{ab}R^{ab}-\frac{d}{4(d-1)}\, R^2\right)
	\right]
	\eea
which is nothing but the higher dimensional extension of NMG \cite{Nakasone:2009bn} . For $K=2$ we get a higher derivative extension of Einstein gravity involving quartic powers of curvature. Explicitly, we have
	\bea
	&& \frac{1}{16}\,\frac{\partial \mathcal L^{(K)}}{\partial g^{ab}}\frac{\partial \mathcal L^{(K)}}{\partial g_{ab}}=
	R^2 R_{a b} R^{a b} - 4\, R\, R_{a b} R^{a}_{c} R_b^c - 4\, R R_{a b} R_{c d} R^{a c b d} +\nonumber \\
	&&	2\, R R^a_b R_{a c d e} R^{b c d e} + 4\, R_a^b R^{a c} R_{b d} R_c^d + 8\, R_{a b} R^a_c R^{d e} R_{b d c e} - 4\, R^{a b} R_{a c} R_{b d e f} R^{c d e f} +\nonumber \\
	&& 4\, R^{a b} R_{c d} R_{a e b f} R^{c e d f} + 4\, R^{a b} R^c_{\ a b d} R_{c e f g} R^{d e f g} + R_{a b c d} R^{a b c e} R^{d f g h} R_{e f g h}
	\eea
We have checked that such a theory satisfies a holographic $c$-theorem, as expected from the general arguments in section \reef{limit} - conformally flat metrics have two derivative equations in these theories. Also, as follows from general arguments in \cite{Oliva:2010zd,Oliva:2010eb}, we can get a quasitopological gravity theory \cite{Myers:2010jv,Myers:2010ru} by adding to the Lagrangian above some combination of Weyl contractions.


\bibliography{Biblio}{}
\bibliographystyle{JHEP}
\end{document}